# Using asymmetric band analysis to interpret the electronic spectroscopy of the Olivine family

Adrian Brown*

*Plancius Research,
Severna Park MD 21146 LLC*

\* Corresponding author

**Abstract**

This paper discusses the mathematical aspects of band fitting and introduces the mathematics of the Asymmetric Gaussian shape and its tangent space for the first time. First, we derive an equation for an Asymmetric Gaussian shape. We then use this equation to derive a rule for the resolution of two Gaussian shaped bands. We then use the Asymmetrical Gaussian equation to derive a Master Equation to fit two overlapping bands. We identify regions of the fitting space where the Asymmetric Gaussian fit is optimal, sub optimal and not optimal. We then demonstrate the use of the Asymmetric Gaussian to fit four overlapping Gaussian bands, and show how this is relevant to the olivine spectral complex at 1 micron. We develop a new model of the olivine family spectrum based on previous work by Runciman and Burns. The limitations of the asymmetric band fitting method and a critical assessment of three commonly used numerical minimization methods are also provided.

**Introduction**

This paper addresses three mathematical aspects of band fitting: 1.) the standard equations of Asymmetric Gaussian shapes and applications of the tangent or derivative space of the Asymmetric Gaussian shapes and 2.) resolution of two overlapping Gaussian shapes and fitting with a single Asymmetric Gaussian and 3.) estimation of differences for an extreme case of four overlapping bands, which is required for the case of the Olivine family in the visible and near infrared region.

*Causes of Asymmetrical spectral bands*. The Franck-Condon principle states that the most probable kinetic state of an electron when it transfers between quantum states is to retain its prior kinetic energy and direction[1]. In its most perfect realisation, this results in a delta function distribution for band transitions. However this principle can be broken in several directions by conditions in the real world of observational spectroscopy. For example:

1.) Under low pressures, collisions between molecules result in symmetric bell-shaped Gaussian absorption curves whose width depends on the density of the gas. Under higher pressure, absorption bands will broaden due to adiabatic collisions[2]. In wavelength space, this Gaussian band will become asymmetric.

2.) A common cause of asymmetry in Gaussian bands is the lack of instrumental resolution to resolve two nearby bands. This effect will occur for any spectrometer attempting to measure Gaussian bands that partially overlap, but with too few spectral channels across the band to resolve the two shapes.

3.) A less common cause of asymmetrical bands is the case of a spin-forbidden band borrowing energy from a spin-allowed that it is crossing. The combination of the two will create an asymmetric shape that varies with the difference in energy between the bands[3].

The result of all these processes is an asymmetric Gaussian curve, and the methods discussed in this paper are appropriate for determining further information vis-a-vis said asymmetric band. We will also discuss the limiting conditions of when two shapes can be resolved, and what happens when they cannot.

The original understanding of atomic spectroscopy was attained using a Group theoretic[4] approach; its success motivated the search for a deeper understanding of molecular spectra, which introduced a further degree of complexity over atomic spectra[5], as might be expected. Finally, measurement of reflectance spectra of crystals led to the development of crystal field theory which was used successfully on transition metal (3d) spectra. Again, group theory and ligand theory are required for these spectra to be interpreted, including point groups, space groups to understand the optical properties of crystals in the visible and near infrared. Additional selection rules such as the spin selection, Laporte rule (differing parity) and the Jahn-Teller effect[6] all play a part in determining the colour of the transition metal bearing rocks.

Figure 1 presents a crystal model for the Forsterite (Mg) endmember of the olivine family. In this paper, we will use this general model and the Fayalite (Fe) endmember of the family where all the Mg atoms are replaced by Fe. We do not deal with questions of preferred location of cation in this paper, but the reader is referred to many prior research studies that do[7–9].

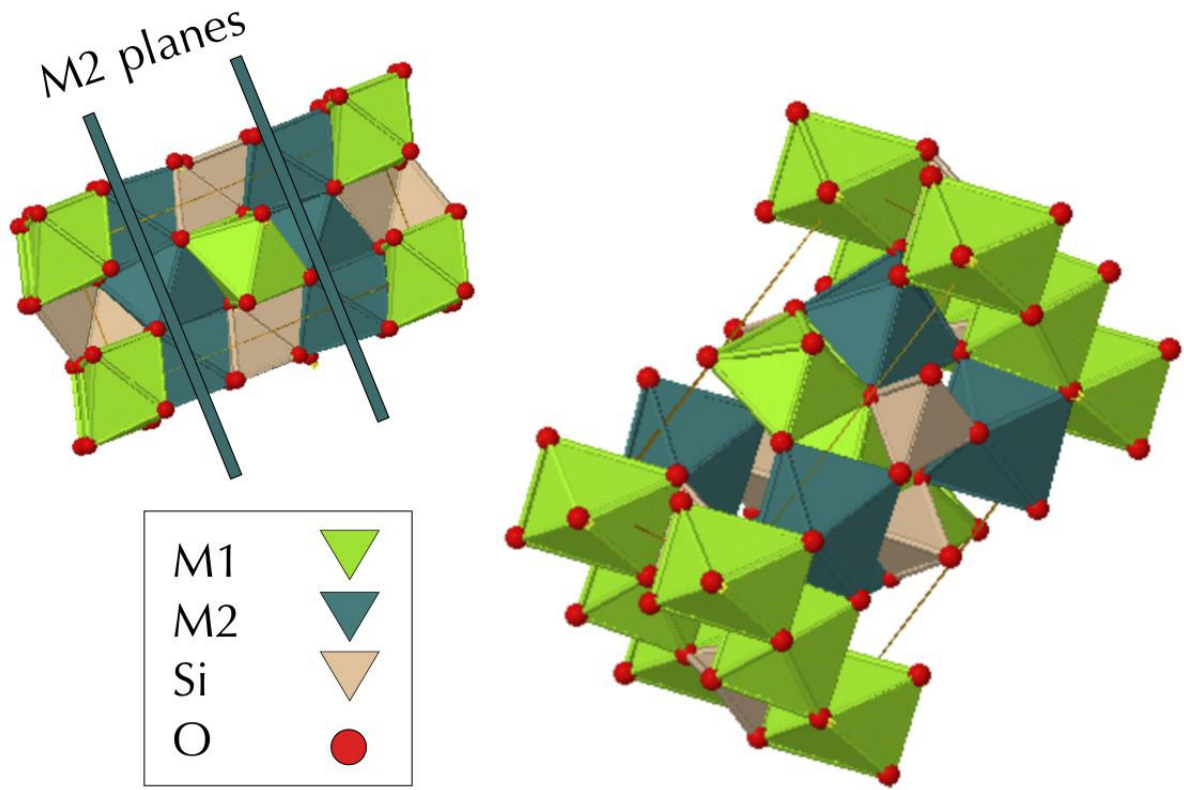

Figure 1 - Olivine family structure showing M1 and M2 octahedral sites, Si tetrahedral sites and M2 planes.

*Rationale for Asymmetric Gaussian analysis*. An easily appreciated rationale for favouring an Asymmetric Gaussian shape over multiple Gaussian shapes[10], is the parameter count. For the Asymmetric Gaussian shapes discussed here, the parameter count is constant at 4 (per shape). For a standard run of the Modified Gaussian Method (MGM), which uses multiple Gaussian curves, the parameter count is 3 per curve, or 9 when fitting the olivine family band complex at 1 micron with 3 standard Gaussian shapes. This is more than double the Asymmetric Gaussian parameter count. For the inclusion of each Gaussian band in a fitting routine, the count scales linearly as $3n$. The benefits of being able to encapsulate or collapse the results of the fitting into one extra parameter is obvious. It even makes possible the analysis of entire hyperspectral images from drill cores[11] or planetary orbital datasets[12,13].

However, two reasonable questions then arise - 1.) How far apart do Gaussian bands have to be to resolve overlapping Gaussian bands from one another? and 2.) Are there differences for the endmembers of the olivine family in the asymmetric shape? This paper uses a mathematical approach to tackle these questions - How much is sacrificed for the gain of using a single parameter in hyperspectral band fitting? It is readily acknowledged that this is only a start to this assessment process.

**Standard Asymmetric Gaussian shape**

Equation (1) gives a standard Gaussian shape that for example has been used to map hyperspectral absorption bands in previous work[14].

$$f(\lambda) = \alpha \exp(-\left[\frac{\lambda - \lambda_0}{\sigma}\right]^2) \tag{1}$$

The three parameters of this equation are the height, $\alpha$, the half width at half maximum, $\sigma$, the centroid, $\lambda_0$. The area under the standard Gaussian curve is $\alpha\sigma\sqrt{\pi}$.

The Asymmetric Gaussian shape as proposed in a previous study[15] is given as the function $g(\lambda)$:

if $\lambda_0 < \lambda$,
$g(\lambda) = \alpha \exp(-\left[\frac{\lambda - \lambda_0}{\sigma}\right]^2)$
else if $\lambda_0 \geq \lambda$, $\tag{2}$
$g(\lambda) = \alpha \exp(-\left[\frac{\lambda - \lambda_0}{\chi\sigma}\right]^2)$

The Asymmetric Gaussian shape is identical to the Gaussian shape, but for the addition of the asymmetry parameter $\chi$ which is inserted on one side as a multiplier of the denominator of the exponent. Effectively, this parameter is unbounded, and can have any value on the positive real number line $\mathbb{R}^+$. When $\chi$ is 1, we have a standard Gaussian shape. When it is less than 1, the shape has right asymmetry. When $\chi$ is greater than 1, the shape possesses left asymmetry (Figure 2).

To further examine the effect of the asymmetry parameter $\chi$, we shall calculate the half width half maximum (HWHM) of the Asymmetric Gaussian shape.

We wish to find the solution to equation (2) at the half maximum value, where $g(\lambda) = \alpha/2$:
$\alpha/2 = \alpha \exp(-\left[\frac{\lambda - \lambda_0}{\chi\sigma}\right]^2)$
Dividing out $\alpha$ and taking $ln$ on both sides:
$-ln(2) = -\left[\frac{\lambda - \lambda_0}{\chi\sigma}\right]^2$

$ln(2) = \left[\frac{\lambda - \lambda_0}{\chi\sigma}\right]^2$
$ln(2)(\chi\sigma)^2 = \left[\lambda - \lambda_0\right]^2$
$\lambda = +/- \chi\sigma\sqrt{ln(2)} - \lambda_0$
So calling the left and right $\lambda$ values $\lambda_a$ and $\lambda_b$ and subtracting them from each other, we obtain:
$HWHM = \lambda_a - \lambda_b = 2\chi\sigma\sqrt{ln(2)} \approx 1.6651\chi\sigma$

Therefore the effect of the asymmetry parameter is to change the width of the band to the same order as the width parameter $\sigma$.

For the Gaussian side of the shape, the asymmetry parameter is 1.

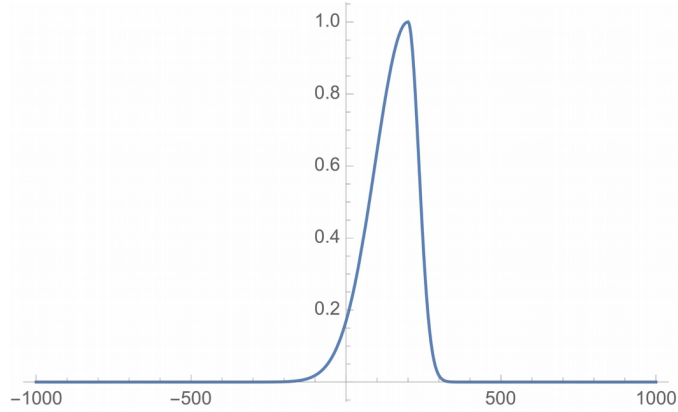

Figure 2 - Asymmetric Gaussian shape showing left asymmetry. Eqn (2) was used to construct the curve, parameters are $\lambda_0$=200, ?=1, ? =50, ?=3.

**Resolution of Gaussians using the Tangent approach**

The derivative of a standard Gaussian shape is $f'$:

$$f' = \frac{df}{d\lambda} = -\frac{2(\lambda - \lambda_0)}{\sigma^2} \alpha \exp(-\left[\frac{\lambda-\lambda_0}{\sigma}\right]^2) \qquad (3)$$

Therefore the derivative of the Asymmetric Gaussian shape is $g'$:

if $\lambda_0 < \lambda$,
$$g' = \frac{dg}{d\lambda} = -\frac{2(\lambda - \lambda_0)}{\sigma^2} \alpha \exp(-\left[\frac{\lambda-\lambda_0}{\sigma}\right]^2)$$
else if $\lambda_0 \geq \lambda$
$$g' = \frac{dg}{d\lambda} = -\frac{2(\lambda - \lambda_0)}{(\chi\sigma)^2} \alpha \exp(-\left[\frac{\lambda-\lambda_0}{\chi\sigma}\right]^2) \qquad (4)$$

Figure 3a demonstrates the relative difficulty of resolving two Gaussian shapes from each other, as their centroids move apart. To construct the figure, we have used what we call the tangent approach to determining the resolvability of two Gaussian shapes. We have used the values of equation (4) at the centroid of the smaller Gaussian shape as a measure of resolvability.
This metric works because when the two shapes are completely separate from each other, the tangent value should be zero. As the two shapes interfere, the value of the tangent increases.

The tangent of the Gaussian shape is given in equation (3). This shows that as the half width grows, the tangent will decrease. As the amplitude grows, the tangent grows.

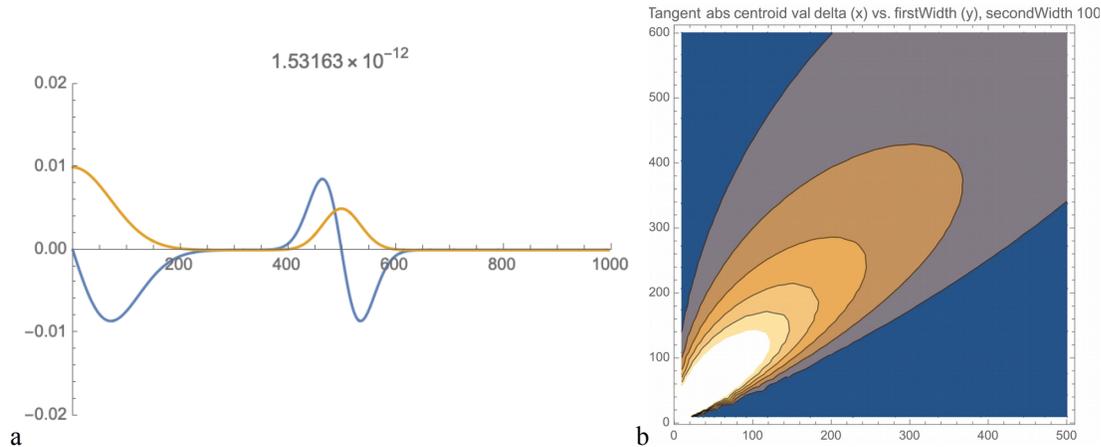

a
b

Figure 3 - Resolution of Gaussian shapes using the tangent approach. (a) .gif movie of the tangent space reaction to a Gaussian with centroid running from 1 to 900. The other Gaussian at 500. (b) This plot shows the variation of the tangent metric with the delta (difference between centroids) and the width of the larger Gaussian shape.

Figure 3b demonstrates several aspects of the resolution of the Gaussians problem. In a first order analysis, the figure shows a large diagonal maximum across the figure, which demonstrates that when the difference is similar to the width, the resolution is difficult.

Second order effects are just as important and highlight the interesting nature of this separation problem. First, there is an area of low values in the top left corner of the plot, indicating that for low delta values, the Gaussians are still resolvable, *if* the width of the larger Gaussian is large enough.

In addition, and the final aspect of the plot we will discuss, is that the broad maximum in the middle of the plot decreases as the width and the delta increase. The interpretation of this observation is that if the width and delta are similar, it is generally harder to resolve the Gaussians for lower delta and width, and easier to resolve them as they increase together.

**Multiple Gaussian curves fit with a single Asymmetric Gaussian**

In this section is provided a simulation of multiple Gaussian shapes being fit by a single Asymmetric Gaussian curve, in order to provide an appreciation for the Asymmetric band fitting method.

Let's begin with the case of two Gaussian curves being fit by one Asymmetric Gaussian. In Figure 3a is plotted a test spectrum of two Gaussian functions in energy space and the finite difference of this curve. Also plotted in Figure 3a is the tangent space obtained using finite derivatives.

The situation in Figure 3a can be represented by the following equation for the sum of two overlapping Gaussian shapes:

$$f(\lambda) = \alpha_0 \exp\left(-\left[\tfrac{\lambda-\lambda_0}{\sigma_0}\right]^2\right) + \alpha_1 \exp\left(-\left[\tfrac{\lambda-\lambda_1}{\sigma_1}\right]^2\right) \tag{6}$$

Subtracting the Asymmetric shape in equation (2), we obtain our *Master Equation* for this study:

$$\alpha_0 \exp\left(-\left[\tfrac{\lambda-\lambda_0}{\sigma_0}\right]^2\right) + \alpha_1 \exp\left(-\left[\tfrac{\lambda-\lambda_1}{\sigma_1}\right]^2\right) - \alpha_2 \exp\left(-\left[\tfrac{\lambda-\lambda_2}{\chi\sigma}\right]^2\right) \tag{7}$$

where if $\lambda_2 < \lambda$, $\chi$=1. We can then explore the stability of equation (7) by using a minimization scheme to best approximate the four parameters of the equation.

*Stability and accuracy of the fitting scheme for two Gaussians.* Let us pause for a moment to discuss the stability and our choice of fitting scheme. Figure 4 shows the results of a least squares fitting approach to our Master Equation (7). These three images show the presence of strong minima that demonstrates the stability of the minimization scheme. To obtain these plots, we hold the parameters of the two Gaussian shapes constant, and vary the Asymmetric Gaussian shape and record the value of our Master Equation. Table 1 displays the parameter space we have explored to create the plots in Figure 4. Essentially we have fixed the amplitude so one Gaussian shape is twice as high as the other and fixed the width of the smaller Gaussian to be half that of the taller Gaussian. We have fixed the difference between the two centroids ($\lambda_1$-$\lambda_0$) to be 100. We call these the Target parameters in the Table.

We then ran a minimization routine using a Levenberg-Marquardt routine to find the optimal parameters for the Asymmetric Gaussian shape, and held these other values constant at those best fit values while each of the three parameters was varied against the centroid to create the contour plots. We call these the Best fit asymmetric parameters in the Table.

| | | | Best fit asymmetric parameters | | | |
|---|---|---|---|---|---|---|
| | $?_0$/$?_1$ | $\lambda_1$-$\lambda_0$ | ? | $\lambda_2$ | $?_2$ | $?_2$ |
| 2 | 2 | 100 | 0.587 | 428.229 | 238.108 | 1.3817 |

Table 1. Parameter space explored for the two Gaussian minimization problem.

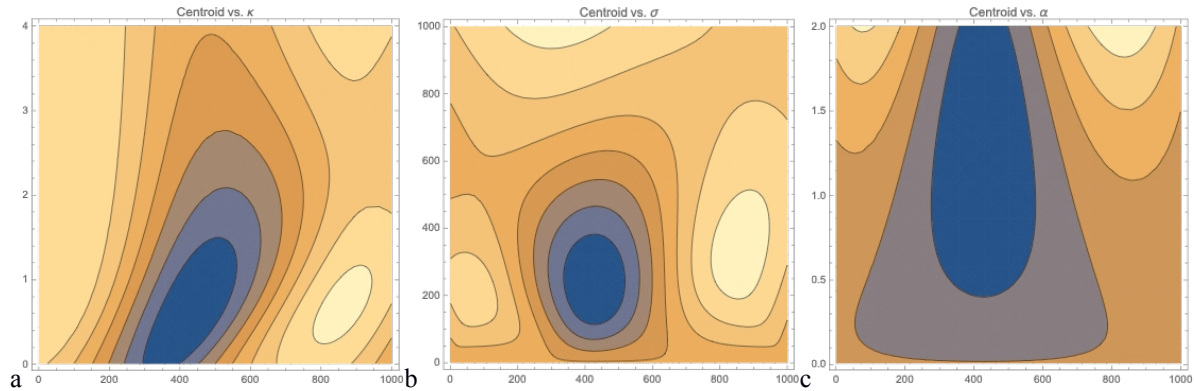

Figure 4. Minimization results for the Master Equation (a) Contour plot showing minimization result for variation of the centroid and asymmetry parameter (c) Contour plot showing variation of centroid versus width (c) Contour plot showing variation of centroid versus amplitude.

Figure 4 is designed to give an assessment of the sensitivity of each of the parameters to the varied parameters, and to give an assessment of the solution landscape. In each of the three plots, the global minimum has been reached, and although there are some local minima visible in surrounding regions.

*State of the fitting scheme*. In a two Gaussian fitting situation, we can imagine a situation in which the curve smaller in amplitude and smaller in width (the opposite situation will have an inverse result) starts at the same centroid as the larger Gaussian, and is pulled away in a series of steps, until they are no longer in contact. This process was shown in Figure 3a. We now wish to examine that situation from the point of view of how suitable an Asymmetric Gaussian approach would be at different stages as two Gaussians are pulled apart from each other.

Figure 5 shows this situation graphically, showing the error of the fitting distance versus the distance between the two centroids of the Gaussians. Table 1 contains the parameters for this run.

We will now split this situation up into three states and discuss each individually.

State 1 -> **Optimal**. In the optimal region, we can see that the minimum error is achieved in the region as wide as the distance between the Gaussian shapes (100 in this case, see Table 1). Up until that point, the minimum error shrinks at a weak rate and is approximately linear. This can be envisaged as the region in Figure 3a where the choice of the asymmetric fit is optimal, the bands are not yet separated and an individual peak of the second Gaussian is not yet apparent. This is the easiest shape to fit with an Asymmetric Gaussian.

State 2-> **Suboptimal**. As the smaller Gaussian continues to peel away from the larger, the state of the fit enters the suboptimal stage. For the situation depicted in Figure 5, this results in an approximately linear ramp from the optimal error, starting at about the width of the larger Gaussian, followed by a flattened region again the same width as the larger Gaussian.

State 3-> **Not Optimal**. As the smaller Gaussian breaks entirely free of the larger, the Not optimal region is entered. At this stage, the fit increases at a higher rate than the previous ramp. In this region, the two Gaussians should be separate Gaussian shapes. The Asymmetric Gaussian approach is then Not Optimal for this situation.

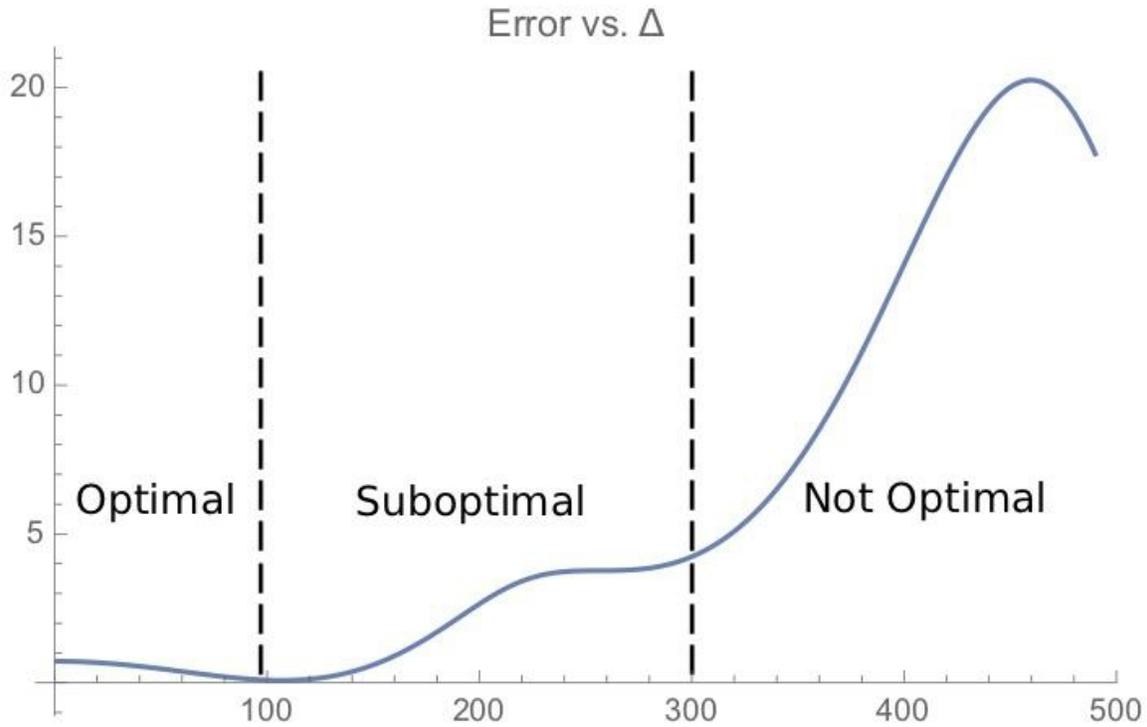

Figure 5. State plot for variations of centroid distance (Δ) showing Optimal, Suboptimal and Not Optimal regions of the field.

Figure 6 demonstrates the error "energy space" plot for three different minimization approaches. Some fitting schemes require a gradient for the next iteration for fast convergence[16]. The tangent space is often put to use for variants on Newton's method for fitting a curve, and we will compare two such methods in this paper to the Nelder-Mead method, which requires no derivatives. The first is the Levenberg[17]-Marquardt[18] method, and second is the Broyden[19] Fletcher[20] Goldfarb[21] Shanno[22] (BFGS) method, which is termed a Quasi-Newton method.

We also point out that we have used the Nelder-Mead approach in past work[13], however we found that it was capable of producing a higher number of singularities when the difference between the centroids $\lambda_1-\lambda_0$ was close to zero. Although this is a challenging fitting case that is not a normal situation, we also found that the BFGS Quasi-Newton scheme was quicker than the Nelder-Mead and Levenberg-Marquadt schemes and fewer singularities appeared in the results, as can be seen from a comparison of Fig. 7 a, b and c.

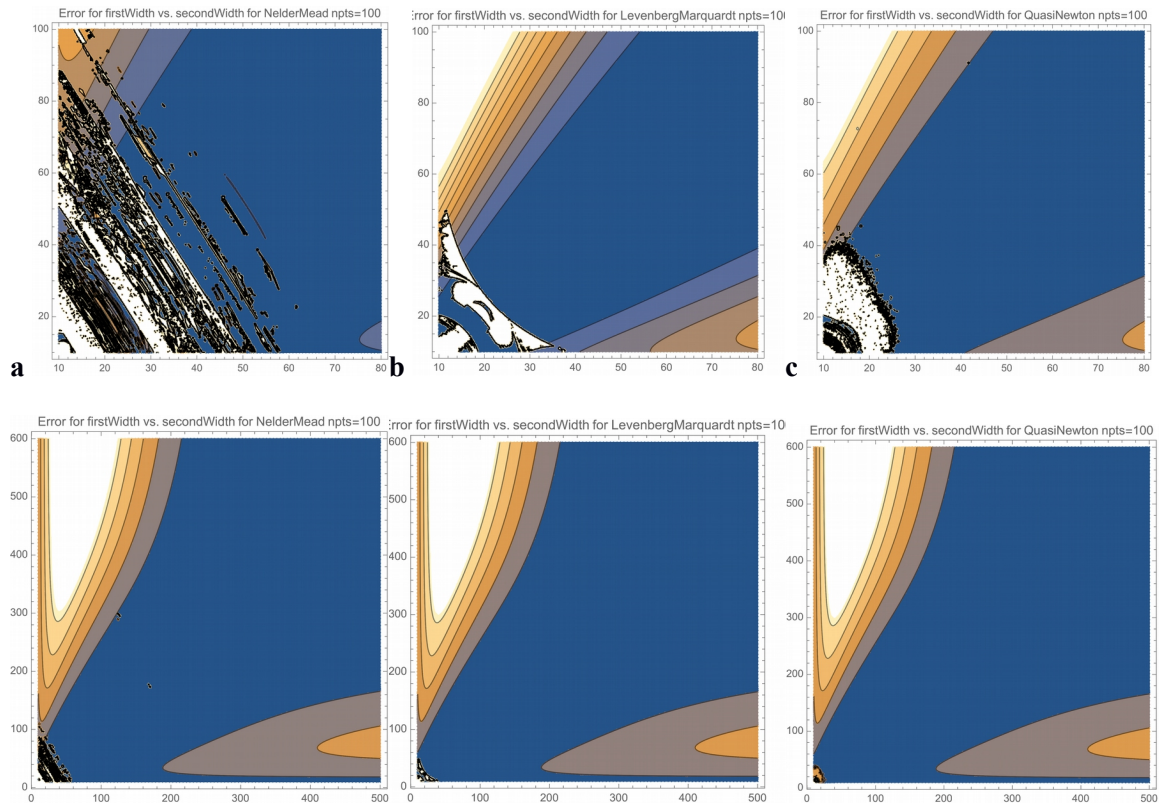

Figure 6. Error plots for three types of fit - a) Nelder Mead, b) Levenberg-Marquardt and c) Quasi Newton. The x-axis is the width of the first Gaussian shape, the y-axis is the width of the second Gaussian shape, which is half as large as the first.

Figure 6 also shows that for all three fitting routines, the errors were slightly more sensitive to the width of the second highest peak than that of the highest peak. This can be seen by the higher errors in the y-axis (second width), compared to lower errors in the x-axis (first width) of Figure 6. The reason for this asymmetry of sensitivity is because the amplitude of the first Gaussian is double that of the second Gaussian (Table 1). If the two shapes were the same height, the asymmetry would disappear.

*Parameter tuning by singularities*. Figure 6 demonstrates graphically the relative performance of the fitting systems. We also used this potential surface to choose the correct initial conditions for the minimization routine – the presence of singularities throughout the potential surface allowed us to fine tune the fit parameters, especially in the case of the Nelder-Mead system which is almost too flexible, and is sensitive to the fitting parameters. These initial conditions were varied until the singularities in the phase space were minimized.

**Olivine 1 micron multiband complex**

Finally, we address the case of the representation of four overlapping Gaussian curves. This system is relevant to the case of the electronic bands of olivine, which overlap in the 1 micron region. Figure 7 shows a possible example of this set up for an olivine electronic transition[7]. As we have previously discussed, we can use information from the tangent space to separate Gaussian bands. This suggests to us that the bands will not be separable if they are closer than the full width half maximum of the largest band.

As previously established by the work of Burns, the Crystal Field Theory applied to the olivine $(Mg^{2+},Fe^{2+})_2SiO_4$ structure gives us four olivine absorption bands in the 1 micron region. Our task is to generate spectra from the following model.

The Olivine family has an orthorhombic space group[23] with a Hermann-Marguin crystallographic *space* group of *Pbnm* (No. 62) and Schoenflies space group of $V^{16}_h$. Olivine's optical constants have previously been derived[24,25] and the electronic structure has been studied using multiple instrumental techniques[26–28]. Of interest to us here is the fact that the transition metals in olivine (here we consider Mg and Fe) are placed in two orthrhombic structural cages called M1 and M2 (Figure 1) and this results in a splitting of the crystal (or ligand) field energy. The *point* group of the M1 site is $C_i$ and the M2 site has $C_s$ symmetry, as first proposed and discussed by Runciman et al[29].

In order to achieve our goal of modeling the 1 micron region of the Mg-Fe olivine family, we use the dataset of Burns et al[30] which provides tables of the centroid of four bands and maps their changes as the Mg-Fe content is varied. Sunshine and Pieters[10] Table 2 showed that the two Y bands in Burns et al[30] have a band width of about 3000cm$^{-1}$ and the Z band has a half width of about 1600cm$^{-1}$. We use these values for our band widths of the Z band and Y bands. Sunshine and Pieters point out they could not resolve the X band and so they ignored that band in their study. Figure 7 shows the X band runs parallel and very close to the bottom of the largest Y band and is very difficult to resolve for this reason if polarized measurements are not available. We use the four band approach here.

Figure 7a shows the error in fitting the four band olivine complex at 1 micron. It demonstrates that the case of Fayalite (Fo10) is more difficult to fit than Forsterite (Fo100). The reason for this behaviour is that the Fayalite spectrum is more asymmetric than the Forsterite spectrum. This fact is born out by simple observations of the two end members which are given in Figure 7b.

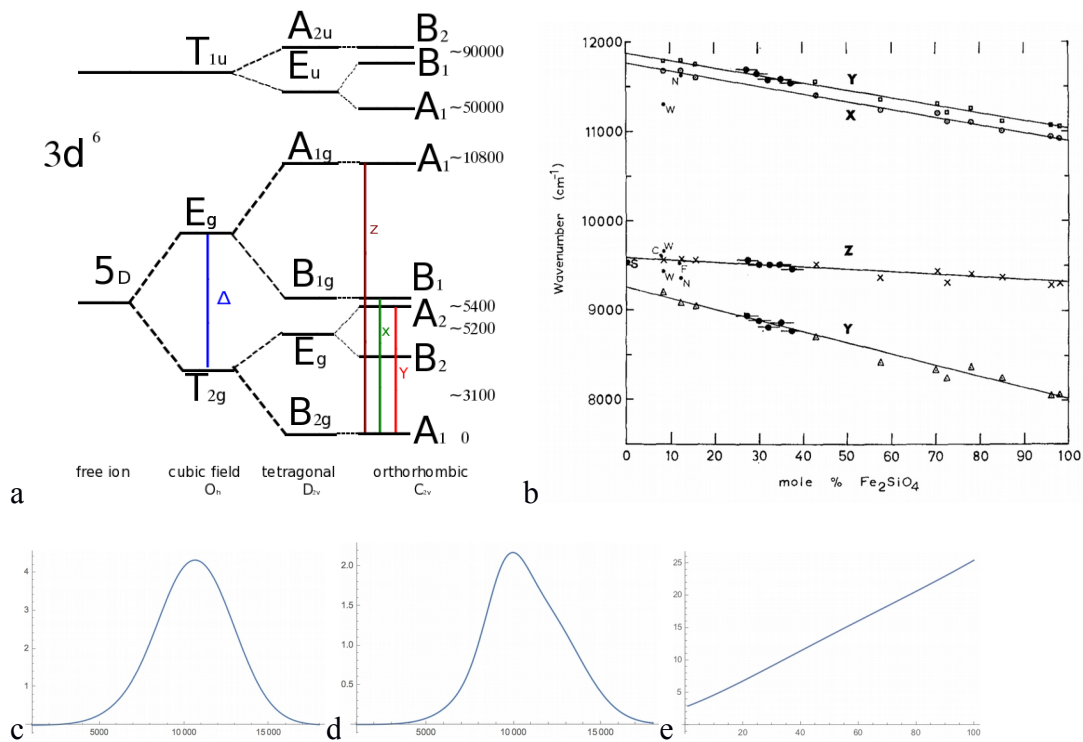

Figure 7(a) Olivine family energy level splitting diagram, modified from Runciman et al.[31] (b) Plot showing the movement of X, Y Y, Z bands between Fayalite and Forsterite endmembers, from Burns et al.[30] (c-d) Example Fo and Fa endmembers (e) Error of fitting of the Fa and Fo endmembers.

## Conclusions

This paper has highlighted the following new findings regarding the fitting of absorption bands with an asymmetric shape:

1.) We demonstrated several aspects of the resolution of the Gaussians problem (Figure 3). In a first order analysis, the figure shows a large maximum diagonally across the figure, which demonstrates that when the delta is similar to the width, it is difficult to resolve the two Gaussians. A second order effect is that for low delta values, the Gaussians are still resolvable, *if* the width of the larger Gaussian is large enough. Finally we established that if the width and delta are similar, it is more difficult to resolve the Gaussians for lower delta and width, and easier to resolve them as they increase together.

2.) Three local minimization/fitting methods were quantitatively tested for the fitting process (Figure 4-6). Singularities of the fitting process were identified and used as guidance for the initial conditions of the fitting parameters. We found that:

    a.) QuasiNewton achieved the best fits, followed by Levenburg-Marquardt and then Nelder-Mead.
    b.) The QuasiNewton method was fastest, followed by Levenburg-Marquardt and

then Nelder-Mead.

3.) The three fitting methods were found to be stable outside of a narrow range of the widths of the first and second Gaussian shapes. All methods struggled as the width of the band became more like a delta function (Figure 5-6)

4.) We found the fitting routines were slightly more sensitive to the width of the second highest peak than that of the highest peak (Figure 6).

5.) We established for the first time that the Asymmetric Gaussian shape can be used to map a parameterised 4 Gaussian shape based on the olivine 1 micron multiband complex. We demonstrated that it is more difficult to fit a Fayalite (Fa) spectrum than a pure Forsterite spectrum, because the Fo spectrum is more symmetric than a Fa spectrum.

The findings of this paper will be used to inform future studies of the olivine visible and near infrared spectrum, and we have shown the utility of the Asymmetric Gaussian shape to elucidate information on the shape of this olivine complex. Future work will assess the pyroxene 1 and 2 micron bands.

## Acknowledgements

This work was supported by NASA MDAP Grant #NNX16AJ48G. I would like to thank Janice Bishop for her advice and encouragement in this project and Michael Chesnes and John Grunwell of the Goddard SFC library for research assistance that helped this project to completion.